\RequirePackage{rotating}
\documentclass[fleqn,usenatbib]{mnras}

\usepackage{rotating, caption, natbib, aas_macros, rotfloat}
\usepackage{newtxtext,newtxmath}
\usepackage[T1]{fontenc}
\usepackage{ae,aecompl}
\usepackage{graphicx}   
\usepackage{amsmath}    
\usepackage{amssymb}    

\newcommand{\header}[1]{\multicolumn{1}{c|}{\textrm{#1}}}
\newcommand{\headerb}[1]{\multicolumn{1}{|c|}{\textrm{#1}}}
\renewcommand{\ion}[2]{#1~{\sc{\romannumeral #2}}}
\def\refitem#1{\relax}

\journal{ISSN 1063-7729, Astronomy Reports, 2019, Vol. 63, No. 12, pp. 1010--1021 (2019) \hfill MNRAS style}

\title[HFS in VALD]{Hyperfine Splitting in the VALD Database of Spectral-line Parameters}

\author[Yu. Pakhomov et al.]
{Yu.V. Pakhomov$^{1}$\thanks{E-mail: pakhomov@inasan.ru}
T.A. Ryabchikova$^{1}$
N.E. Piskunov$^{2}$
\\
$^{1}$Institute of Astronomy, Russian Academy of Sciences, Moscow, Russia\\
$^{2}$Department of Physics and Astronomy, Uppsala University, Uppsala, Sweden
}

\begin{document}

\maketitle

\pagestyle{myheadings}

\begin{abstract}
The Vienna Atomic Line Database (VALD) has been supplemented with new data and new functionality -- the possibility of taking into account the effect of hyperfine splitting (HFS) of atomic levels in the analysis of line profiles. This has been done through the creation of an ancillary SQL database with the HFS constants for atomic levels of 58 isotopes of 30 neutral and singly-ionized atoms. The completeness of the collected data and new opportunities for studies of stars of various spectral types is analyzed. The database enables analysis of splitting of up to 60\%\ of lines with measurable effects in the ultraviolet ($\lambda\gtrsim1000$~\AA), and up to 100\%\ of such lines in the optical and infrared ranges ($\lambda\lesssim25000$~\AA) for A--M stars. In the spectra of hot O--B stars, it is necessary to use laboratory measurements for atoms in the second and higher stages of ionization.

\end{abstract}

\section{INTRODUCTION}
\noindent

VALD (Vienna Atomic Line Database) is a database of parameters of spectral lines intended for the creation of samples of data needed for the calculation and analysis of stellar spectra~\citep{1995A&AS..112..525P, 1999A&AS..138..119K, 2015PhyS...90e4005R}. In VALD, every spectral line is characterized by an element identifier, its ionization stage, wavelength $\lambda$, transition probability (oscillator strength), and damping factors (radiative, Stark, van der Waals); for the lower and upper levels, the excitation energy $E$, total angular momentum quantum number $J$, Land\'e factors $g$ are given as well, together with a description of the electronic configuration and term designation. Currently, the VALD data enable a fairly precise description of the observed profiles of most spectral lines. However, high resolution spectra contain a number of lines whose profiles are additionally broadened by hyperfine splitting (HFS), whose description requires additional data. Without these data, it is impossible to obtain high-precision values of the abundances of elements such as lithium, manganese, cobalt, copper etc., especially, heavy elements. The uncertainty in the abundance of an element depends on the magnitude of the HFS of the lines and their intensity, and can reach 1 dex or more in some cases~\citep{2008ARep...52..630B}. Thus, in many cases it is imperative to take into account HFS when interpreting spectra.

In the present paper, we describe a supplemental database to VALD enabling calculations of theoretical spectra taking into account HFS. Section~\ref{sec:HFS} examines the data needed to calculate split lines profiles and describes the database of HFS constants. In Section~\ref{sec:completeness}, we analyze the completeness of the data presented. In Section~\ref{sec:dis}, we discuss the possibility of applying this new functionality to VALD.

\section{THE EFFECT OF HYPERFINE SPLITTING}
\label{sec:HFS}
\noindent

\begin{table*}
	\centering
	\caption{ Characteristics of the HFS data}
	\label{tab:elem}
	\renewcommand\arraystretch{1.3}
	\tabcolsep=1.5mm
	\begin{tabular}{|l|c|r|r|r|r|p{5mm}|l|c|r|r|r|r|}
		\cline{1-6}\cline{8-13}
		\headerb{Isotope}  & \header{Numbers}   & \header{$E_{min}$} & \header{$E_{max}$} & \header{$A_{min}$} & \header{$A_{max}$} && 
		\header{Isotope}  & \header{Numbers}   & \header{$E_{min}$} & \header{$E_{max}$} & \header{$A_{min}$} & \header{$A_{max}$}\\
		\headerb{ion}     & \header{of levels} & \header{cm$^{-1}$} & \header{cm$^{-1}$} & \header{MHz} & \header{MHz} &&
		\header{ion}     & \header{of levels} & \header{cm$^{-1}$} & \header{cm$^{-1}$} & \header{MHz} & \header{MHz}  \\
		\cline{1-6}\cline{8-13}
		\ion{$^{6}$Li}{1}    &   3 &      0 &  14903 &     -1 &    152 &&  \ion{$^{71}$Ga}{1}   &   5 &      0 &  38914 &    299 &   7255 \\
		\ion{$^{7}$Li}{1}    &   3 &      0 &  14903 &     -3 &    401 &&  \ion{$^{71}$Ga}{2}   &   8 &  47814 & 145493 &   1439 &   9923 \\
		\ion{$^{23}$Na}{1}   &  60 &      0 &  40888 &      0 &    885 &&  \ion{$^{85}$Rb}{1}   &   3 &      0 &  12816 &     25 &   1011 \\
		\ion{$^{27}$Al}{1}   &  23 &      0 &  46183 &    -99 &    502 &&  \ion{$^{87}$Rb}{1}   &   3 &      0 &  12816 &     84 &   3417 \\
		\ion{$^{27}$Al}{2}   &  44 &      0 & 149182 &   -869 &   2728 &&  \ion{$^{89}$Y}{2}    &  39 &      0 &  84275 &   -241 &    232 \\
		\ion{$^{39}$K}{1}    &  21 &      0 &  31765 &      0 &    230 &&  \ion{$^{93}$Nb}{2}   &  53 &      0 &  40561 &  -1145 &   1229 \\
		\ion{$^{40}$K}{1}    &   5 &      0 &  21536 &   -285 &      1 &&  \ion{$^{95}$Mo}{2}   &  21 &      0 &  59840 &   -919 &    460 \\
		\ion{$^{41}$K}{1}    &   5 &      0 &  21536 &      0 &    127 &&  \ion{$^{97}$Mo}{2}   &  21 &      0 &  59840 &   -940 &    470 \\
		\ion{$^{45}$Sc}{1}   &  45 &      0 &  34422 &   -738 &   2232 &&  \ion{$^{127}$I}{1}   & 107 &  54633 &  82615 &  -1319 &   4742 \\
		\ion{$^{45}$Sc}{2}   &  42 &      0 &  85832 &   -480 &    654 &&  \ion{$^{127}$I}{2}   &  61 &   7087 & 136733 &   -630 &   5294 \\
		\ion{$^{47}$Ti}{1}   &  34 &      0 &  33700 &   -145 &     79 &&  \ion{$^{135}$Ba}{2}  &  38 &      0 &  75945 &    -11 &   3591 \\
		\ion{$^{47}$Ti}{2}   &  80 &      0 &  47625 &   -858 &    174 &&  \ion{$^{137}$Ba}{2}  &  38 &      0 &  75945 &    -11 &   4018 \\
		\ion{$^{49}$Ti}{1}   &  34 &      0 &  33700 &   -145 &     79 &&  \ion{$^{139}$La}{2}  & 108 &      0 &  66591 &  -1128 &   2951 \\
		\ion{$^{49}$Ti}{2}   &  80 &      0 &  47625 &   -858 &    174 &&  \ion{$^{141}$Pr}{1}  &  90 &   6714 &  34190 &   -994 &   1378 \\
		\ion{$^{50}$V}{1}    &  11 &      0 &  18438 &    141 &    356 &&  \ion{$^{141}$Pr}{2}  & 293 &      0 &  42194 &   -273 &   2114 \\
		\ion{$^{51}$V}{1}    & 332 &      0 &  46230 &   -947 &   1445 &&  \ion{$^{151}$Eu}{2}  &  13 &      0 &  27256 &  -1672 &   1540 \\
		\ion{$^{51}$V}{2}    & 116 &      0 &  90584 &   -411 &   1479 &&  \ion{$^{153}$Eu}{2}  &  13 &      0 &  27256 &   -743 &    684 \\
		\ion{$^{55}$Mn}{1}   & 144 &      0 &  61225 &  -2196 &   1848 &&  \ion{$^{155}$Gd}{2}  &  68 &   2856 &  40924 &   -349 &    407 \\
		\ion{$^{55}$Mn}{2}   & 221 &      0 & 119197 &  -1151 &   1604 &&  \ion{$^{157}$Gd}{2}  &  67 &   2856 &  40924 &   -457 &    534 \\
		\ion{$^{57}$Fe}{1}   &  57 &      0 &  61064 &    -40 &    143 &&  \ion{$^{159}$Tb}{2}  &  79 &      0 &  36000 &   -271 &   1700 \\
		\ion{$^{59}$Co}{1}   & 371 &      0 &  60262 &  -1016 &   3084 &&  \ion{$^{165}$Ho}{1}  & 165 &      0 &  42381 &    265 &   1486 \\
		\ion{$^{59}$Co}{2}   &   4 &  17771 &  56010 &    149 &   2398 &&  \ion{$^{175}$Lu}{1}  &  17 &      0 &  40735 &   -924 &   4511 \\
		\ion{$^{61}$Ni}{1}   &   6 &      0 &  29888 &   -457 &    -78 &&  \ion{$^{175}$Lu}{2}  &   4 &      0 &  28503 &  -2038 &   4976 \\
		\ion{$^{63}$Cu}{1}   &  70 &      0 &  71290 &   -510 &   5866 &&  \ion{$^{176}$Lu}{1}  &  19 &      0 &  40735 &   -651 &   3189 \\
		\ion{$^{65}$Cu}{1}   &  70 &      0 &  71290 &   -543 &   6284 &&  \ion{$^{181}$Ta}{1}  & 507 &      0 &  59300 &  -3515 &   7237 \\
		\ion{$^{67}$Zn}{1}   &  13 &      0 &  80175 &      0 &    609 &&  \ion{$^{181}$Ta}{2}  & 210 &      0 &  81164 &  -2348 &   4960 \\
		\ion{$^{67}$Zn}{2}   &   2 &  48481 &  65441 &    357 &    549 &&  \ion{$^{203}$Tl}{1}  &   2 &      0 &   7792 &   1049 &  21111 \\
		\ion{$^{69}$Ga}{1}   &   5 &      0 &  38914 &    239 &   5726 &&  \ion{$^{205}$Tl}{1}  &   4 &      0 &  34159 &   1061 &  21315 \\
		\ion{$^{69}$Ga}{2}   &   8 &  47814 & 145493 &   1109 &   7794 &&  \ion{$^{209}$Bi}{3}  &   5 &      0 &  89236 &    647 &  51050 \\
		\cline{1-6}\cline{8-13}
	\end{tabular} 
\end{table*}

The effect of HFS of the energy levels in an atom is due to the interaction of the magnetic moment of the nucleus with the magnetic field of an electron created by its orbital motion~\citep{Sobelman}. In terms of quantum numbers, taking into account relativistic effects, the total electron angular momentum $J$ interacts with the magnetic moment of the nucleus $I$, and creates the HFS angular momentum $F$ , which is saved and can take the values $F=|J-I| ... |J+I|$. As a result, a level with quantum number $J$ splits into $2(\mathrm{min}(I,J)+1)$ sub-levels, and the spectral line formed by the transition between levels $J_l$ and $J_u$ of various electronic configurations splits into
multiple components, whose number can exceed 20. The Selection rules for the magnetic-dipole interaction (by analogy with the spin--orbit interaction $\Delta\,J_{lu}=0,\pm 1$) allow transitions with $\Delta\,F_{lu}=0,\pm 1$ and forbid the transition $F_l=0 \rightarrow F_u=0$. For transitions $\Delta\,J_{lu}=\pm 2$ of the electric-quadrupole interaction, transitions in the split line $\Delta\,F_{lu}=\pm 2$ are
allowed.

The shift in the energy of a level subject to HFS relative to the unsplit level, characterized by the quantum numbers $I$, $J$, and $F$, is described by the formula: 
\begin{eqnarray}
	\Delta E = \frac{1}{2}AK + B\frac{(3/4)K(K+1)-J(J+1)I(I+1)}{2I(2I-1)J(2J-1)}, \label{eq:dE}
\end{eqnarray}
where $K=F(F+1)-J(J+1)-I(I+1)$, and $A$ and $B$ are the HFS constants. The value $A$ characterizes the magnetic-dipole interaction, and $B$ the electric-quadrupole interaction. The relative intensity of the components is calculated using 6j symbols:
\begin{eqnarray}
	I_{rel}(F_l\rightarrow F_u)=
	\frac{(2F_l+1)(2F_u+1)}{2I+1}\left\{\begin{array}{ccc}
	J_l & F_l & I \\
	F_u & J_u & 1~|~2 \end{array}\right\}^2, \label{eq:dI}
\end{eqnarray}
where the last number is equal to 1 for a magneticdipole transition and to 2 -- for an electric-quadrupole transition. The condition $\sum I_{rel}=1$ is satisfied.

Thus, to calculate the effect of HFS, one needs to know for lower and upper levels of a transition of energy $E$ the quantum numbers $J$, HFS constants $A$ and $B$, and the magnetic moment of the nucleus of the considered isotope $I$. VALD contains only $E$ and $J$. The values of $I$ can be taken, for example, from~\cite{1998JPCRD..27.1275R}. HFS constants are determined in the laboratory and are regularly published in scientific articles. Therefore, our main task was collecting sources of information on HFS measurements. The most complete set of papers on such measurements is contained in the compilation by R.L.~Kurucz\footnote{http://kurucz.harvard.edu/atoms}, where ab*.dat files containing data for a given isotope and the quantities $E$, $J$, $A$ and $B$ for various levels are presented for some isotopes, as well as bibliographic references. However, some significant elements, such as lithium, praseodymium etc., are absent from this compilation. Therefore, we used the SAO/NASA Astrophysics Data System (ADS) to search for articles containing data on HFS measurements. Due to the limited capabilities of laboratory measurements, HFS data exist only for neutral and singly-ionized atoms whose lines are observed in the near-UV ($\lambda\gtrsim 700$~\AA), optical, and IR ($\lambda\lesssim 10^6$~\AA).

The accumulated data were systematized in the SQL database, the element, ionization stage, isotope identifier in VALD (species ID), element number, isotope mass, nuclear magnetic moment $I$, level energy $E$, total angular momentum quantum number $J$, HFS constants $A$ and $B$, and their errors are presented for each level, together with a key to the bibliographic source included in the VALD bibliographic database in BibTeX format.

Altogether, the database contains information for 3970 levels of 58 isotopes of 30 elements. About half the isotopes belong to neutral atoms, and the other half to singly-ionized atoms. Only for Bismuth (Bi) are there data for the lines of a doubly-ionized atom. The list of isotopes is presented in Table~\ref{tab:elem}, which gives for each the number of levels in the database, the minimum $E_{min}$ and maximum $E_{max}$ energies and minimum and maximum values of the constant $A$, mainly determining the degree of the HFS. 

If a VALD ''select stellar`` request (fetching spectral lines contributing to the spectrum of a star with specific atmospheric parameters) is made with the option to include HFS effects, for each selected line, a check is performed for the presence of HFS data for the lower and upper levels, identified based on the values of $E$ and $J$ for the isotope in question. If there are no data for at least one level, the line is characterized as unsplit. Next, the shifts of the energy levels $\Delta E_l$ and $\Delta E_u$ from relation (\ref{eq:dE}), possible transitions $F_l\rightarrow F_u$ according to the selection rules, the wavelength shifts $\Delta\lambda^{F_l\rightarrow F_u}=(\Delta E_l - \Delta E_u)\lambda^2/c$ for each component, and their relative intensities $I_{rel}^{F_l\rightarrow F_u}$ from relation (\ref{eq:dI}) are calculated. The oscillator strength of the unsplit line log\,$gf$ for each component formed by the transition $F_l\rightarrow F_u$, is replaced by the value log\,$gf$+log\,$I_{rel}^{F_l\rightarrow F_u}$, and the wavelength by $\lambda+\Delta\lambda^{F_l\rightarrow F_u}$. The remaining parameters of the spectral line are unchanged. Instead of a single line, a set of spectral lines determied by the number of HFS components appears in the final output. Since HFS components have a distribution of values of $\Delta\lambda^{F_l\rightarrow F_u}$, the final line list is sorted in order of increasing wavelength and is then issued to the user.

\section{COMPLETENESS OF THE DATA}
\label{sec:completeness}
\noindent

\begin{figure}
	\includegraphics[width=\columnwidth,clip]{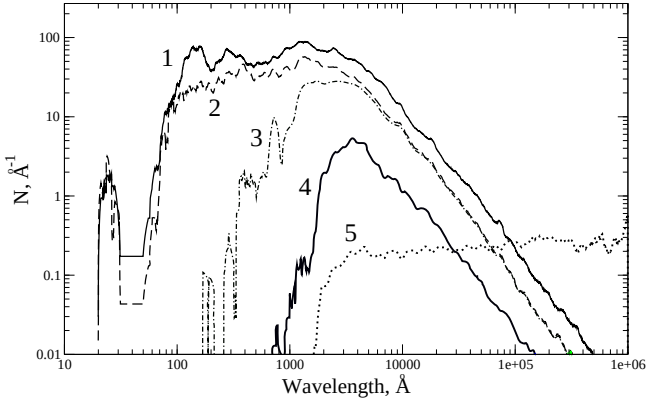}
	\caption{Distribution of the number of spectral lines in VALD as a function of wavelength. Top to bottom are shown: (1) all VALD data, (2) lines with nonzero nuclear magnetic moments, (3) same as 2 for neutral and singly-ionized atoms, (4) lines for which there are HFS data in VALD, and (5) the ratio of 4 and 3.}
	\label{fig:all}
\end{figure}

Figure~\ref{fig:all} shows the distribution of the density of spectral lines (the number of lines per 1~\AA) over the entire wavelength range available in VALD, from 10 to $10^6$~\AA. The uppermost line (1) represents the data for all lines of all elements. A sharp increase in the number of spectral lines is seen near $\lambda$500~\AA. The maximum density of lines reaches about 100 lines per 1~\AA\ in the UV and optical. In the IR range, the density gradually decreases. Approximately every third or fourth line belongs to elements with non-zero nuclear magnetic moments, that is, it is subject to HFS effects (line 2). Line (3) is similar to the previous line, but displays data only for spectral lines belonging to neutral and singly-ionized atoms (there is a dip in the UV due to the large contribution of highly ionized atoms). Ideally, we would like to describe the splitting of all these lines. However, we are limited by existing laboratory data, and in reality we can describe only 10--20\%\ of the spectral lines (5) for $\lambda$>1100~\AA. The restriction on the wavelengths is probably related to difficulties in laboratory measurements in the UV. In the far-IR, the fraction of lines that are subject to HFS reaches 30\%, but their number drops by two orders of magnitude.

In the optical and near-IR, the accumulated data may be useful for $\sim$20\%\ of all spectral lines. This small percentage is due to the fact that VALD contains data for a large number of even fairly weak lines. Laboratory measurements ($\lambda$, $E$, log\,$gf$) are fairly reliable for such lines, but measurements of HFS are complicated, since many components needed to determine the quantum numbers $F$ and splitting constants may be not visible. In practical stellar spectroscopy, most of the missing weak lines do not play a significant role. In this regard, to assess the completeness of the data, we must consider real spectral lines observed in stellar spectra and ones used for their analysis.

\begin{table}
	\centering
	\renewcommand\arraystretch{1.2}
	\caption{ Quantitative characteristics of spectra of stars of various spectral types}
	\label{tab:param}
	\tabcolsep=1mm
	\begin{tabular}{|c|c|c|c|}
		\hline 
		SpType & $T_{eff}$, K/log~$g$ & \multicolumn{2}{c|}{number of selected/I, II/split lines} \\
		\cline{3-4}
		&& central depth $>0.015$ & central depth $>0.30$ \\
		\hline 
		O & 30000/4.0 & 14796/ 265/   1 &  2701/ 245/   0 \\
		B & 20000/4.0 & 18054/ 797/  41 &  4787/ 775/  11 \\
		A &  9500/4.5 & 20997/3728/ 406 &  7064/1065/ 100 \\
		F &  7500/4.5 & 27541/5888/1464 &  9663/1803/ 381 \\
		G &  6500/4.5 & 29980/6725/2082 & 11343/2279/ 671 \\
		K &  4750/4.5 & 33224/8242/3322 & 14045/3005/1436 \\
		M &  3500/4.5 & 26434/7148/2985 & 11514/2810/1415 \\
		\hline 
	\end{tabular} 
\end{table}

\begin{figure}
	\centering
	\includegraphics[width=0.5\textwidth,clip]{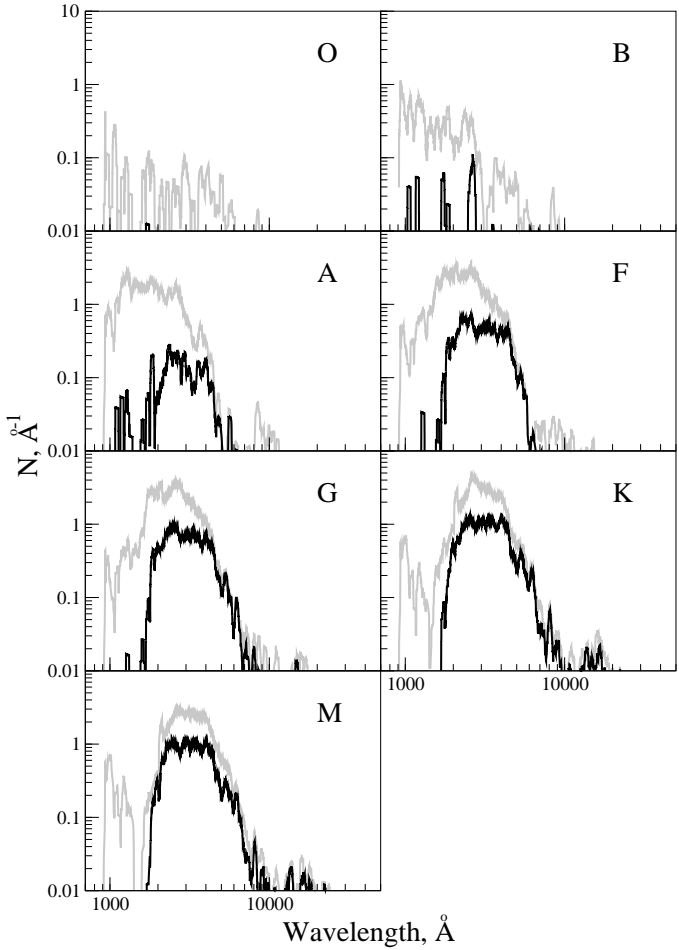}\renewcommand{\baselinestretch}{1}
	\caption{Distribution of spectral lines with central depths exceeding 0.015 in the spectra of stars of various spectral types
		according to VALD. Data for neutral and singly-ionized atoms of elements with nonzero magnetic momentum of the nucleus
		are shown in gray, and lines for which VALD contains HFS constants in black.}
	\label{fig:stars_all}
\end{figure}

Similar plots for main-sequence stars of various spectral types are presented in Fig.~\ref{fig:stars_all}. Only two lines corresponding to the options (3) and (4) from Fig.~\ref{fig:all} are shown, that is, the density of the total number of spectral lines from VALD formed by neutral atoms and single-ionized elements, subject to HFS and those for which there are HFS data in VALD. Calculations were performed using internal VALD commands matching the external query ''select stellar``, in the wavelength range from 912 to 100 000~\AA, where the total number of atomic lines is 956\,112 (omitting molecular lines). Out of these spectral lines, ones with central depths relative to the continuum exceeding 0.015 were selected. We used this criterion because it provides a sufficient quality of the description of the spectrum while cutting off the output for a large number of weak lines that could be selected due to the approximate solution of the transfer equation.

The parameters of the atmosphere ($T_{eff}$/log~$g$) and the number of spectral lines for two sampling criteria (central depth $<0.015$ and $<0.30$) for stars of various spectral types are presented in Table~\ref{tab:param}. The data are represented by three numbers: the total number of selected lines, the number of neutral and singly-ionized atoms of elements with nonzero nucear magnetic moments, and the number of lines whose splitting can be calculated using our database. Figure~\ref{fig:stars_all} shows an appreciable decrease in the amount of data on HFS in the UV. The reason for this is the same as for the dip in Fig.~\ref{fig:all} -- the difficulty of carrying out laboratory measurements in the UV, as well as the fact that lines of atoms and ions that are absent from our database dominate at wavelengths up to 1200~\AA. In UV, the fraction of split lines ranges from 15--20\%\ for hot stars to 50--60\%\ for cool ones. On the contrary, the optical is well studied under laboratory conditions. The fraction of split lines exceeds 50\%\ and reaches 95\%.

\begin{figure}
	\centering
	\includegraphics[width=0.5\textwidth,clip]{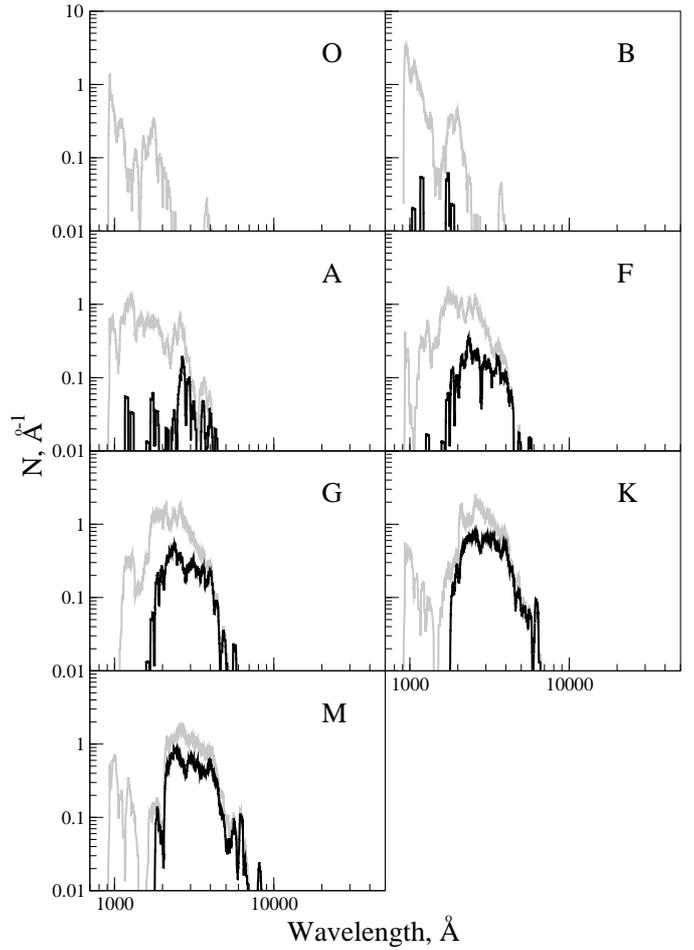}
	\caption{Same as Fig.~\ref{fig:stars_all} but only for lines with computed central depths exceeding 0.3.}
	\label{fig:stars_3}
\end{figure}

The statistics under consideration include all lines in the spectrum with central depths exceeding 0.015 relative to the continuum. However, when analyzing the profiles of individual lines, accounting for HFS of very weak lines does not make practical sense. Therefore, we further consider only lines with theoretical depths exceeding 0.3 relative to the continuum. The real line depth after convolution with the profiles of the macroturbulent velocity, stellar rotation, and instrumental profile will be even lower. For example, for $V_{macro}$=4~km\,s$^{-1}$, $V_{rot}$=1~km\,s$^{-1}$, and $R=80\,000$, the central line depth decreases by about a factor of two. The analysis of the profile of such a line for typical noise levels of observational data already makes sense. The numbers of such lines in the spectra of stars of various spectral types are given in the last column of Table~\ref{tab:param}, which shows that weak lines dominate in terms of numbers, especially for hot stars. Figure~\ref{fig:stars_3} shows results similar to Fig.~\ref{fig:stars_all} obtained for lines with central depths exceeding 0.3. A significant difference is seen for hot stars. There are no lines that could be split using the accumulated data in O stars. The reason for this is simple: lines of elements in high ionization stages, for which no data are available in our database due to the lack of laboratory HFS measurements, dominate in hot stars, and lines of singly-ionized atoms are weak. With decreasing effective temperature of the stars, the number of lines of atoms and first ions begins to increase. In A stars, 40--60\%\ of lines in the UV can be described, and this percentage is almost 100\%\ in the optical. For cooler stars, the percentage of significant split lines is the usual (70--100\%). 

Statistics for the elements are presented in Tables~\ref{tab:nelem}, where the numbers of observed and split lines with central depths $>0.015$ and $>0.3$ for various isotopes in the spectra of stars of various spectral types are listed. The most complete HFS data are available for the isotopes \ion{$^{23}$Na}{1}, \ion{$^{45}$Sc}{2}, and \ion{$^{59}$Co}{1}, as well as \ion{$^{63}$Cu}{1}, \ion{$^{89}$Y}{2}, \ion{$^{139}$La}{2}, \ion{$^{141}$Pr}{2}, \ion{$^{151}$Eu}{2}, and \ion{$^{153}$Eu}{2}, for which we can describe almost the entire set of observed lines. \ion{$^{59}$Co}{1}, the main isotope of cobalt, whose spectrum has been well measured in laboratories, especially stands out. The maximum number of lines of this isotope observed in K stars is 1463, 1457 of which can be split with the new VALD tool.

\section{DISCUSSION}
\label{sec:dis}
\noindent

The database of HFS constants contains data for the main isotopes for which lines are observed in the spectra of stars of various spectral types, as well as data that may be needed in the future for studies of the spectra of plasma under various conditions. It follows from Fig.~\ref{fig:stars_3} that, for A--M stars, the collected data enable us to describe the splitting of 70--100\% of lines in the optical. The data for hotter stars are insufficient, since lines of elements in the second and higher ionization stages dominate in their spectra, for which there are very few laboratory measurements of the splitting parameters. However, the overall ability to describe the spectra of hot stars does not suffer because of this, since the main elements forming lines (up to 90\%) are not affected by HFS.

The database does not contain data for hydrogen, whose nucleus has a non-zero magnetic moment, since the energy transport in its lines demands a separate computation of the line broadening, which usually takes into account HFS. For a number of isotopes in Table~\ref{tab:nelem}, spectral lines cannot be selected for stars of any spectral type. There are several reasons for this. First, VALD contains data on individual isotopes for only a small number of elements: Li, Ca, Ti, Cu, Ba, and Eu. For the remaining elements, the main parameters in VALD are given for a mixture of isotopes. For this reason, HFS data for these latter elements are given only if the most common isotope is subject to HFS. One example is the isotope \ion{$^{57}$Fe}{1}, for which we have HFS data, however, this isotope comprises only 0.021 of all stable isotopes of iron, and the main isotope \ion{$^{56}$Fe}{1} is not subject to HFS. With the further accumulation of data in VALD, splitting calculations will be performed automatically, without changes to the HFS database. Second, the abundances of some elements, such as iodine, can be too low for the appearance of appreciable spectral lines in the spectra of normal stars. However, if their abundance is enhanced for some reason, they can be seen in stellar spectra.

For lithium, data are available for three levels only (significant HFS is manifested only in the ground lithium level $^2$S$_{1/2}$) of two stable isotopes \ion{$^{6,7}$Li}{1}. This admits a description of all 15 components of the fine and hyperfine structures of the $\lambda$6707~\AA\ resonance line, which is important for studies of the evolution of the chemical composition of stars and galaxies. In the Table~\ref{tab:nelem}, the lines of lithium have been selected only for the spectra of cool stars, however, there are other stars whose spectra contain very strong lithium lines, such as young stars and variable stars with active chromospheres (e.g., BY~Dra, RS~CVn stars). Allowance for HFS for lines of sodium, aluminum, and potassium is necessary only under conditions for which the thermal or turbulent velocities do not exceed 1~km/s. For example, the splitting of the D$_{1,2}$ resonance doublet is only 0.02~\AA, and it is even smaller for other lines. HFS increases with increasing magnetic moment, and partially with increasing nuclei mass (see Tab.~\ref{tab:elem}). For some lines of scandium, it is must already be taken into account in the analysis of stellar spectra. Some iron-group elements whose lines form almost entire spectra of certain stars exhibit significant splitting of their levels, and changes in the shapes of their spectral lines are sometimes appreciable even with medium spectral resolution ($R\sim20000$). This also applies to vanadium, manganese, and cobalt. In any case, accurate determination of the abundances of these elements (accuracy not less than 0.10 dex), as well as for copper, requires allowance for HFS.

\begin{figure}
	\centering
	\includegraphics[width=0.5\textwidth,clip]{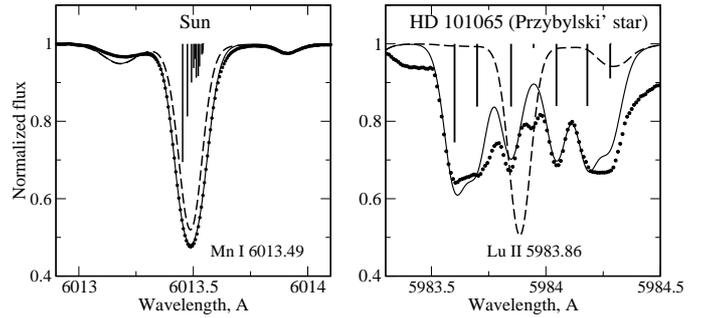}
	\caption{Example of the application of the HFS data. The left panel shows the line  \ion{Mn}{1} 6013.49~\AA\ in the solar spectrum. The
		right panel shows the line \ion{Lu}{2} 5983.86~\AA\ in the spectrum of the star HD~101065 (Przybylski's star) with anomalous elemental
		abundances. The vertical lines show the contributions of particular components.}
	\label{fig:hfs}
\end{figure}

Figure~\ref{fig:hfs} shows the effect of HFS on observed spectral line profiles. The left panel shows the profile of the Mn~I~$\lambda$6013.49~\AA\, line in the solar spectrum. HFS does not strongly change the profile, but it affects the intensity quite substantially. If HFS is not be taken into account, it is necessary to increase the abundance of Mn in the atmosphere by 0.5~dex in order to reproduce the observed profile, which is beautifully described by the modern solar abundance of Mn if HFS is included (black solid curve in Fig.~\ref{fig:hfs}).

The HFS data are crucial for describing the line profiles of rare-earth and heavy elements. Their strong splitting means that even individual HFS components are sometimes visible in high-resolution stellar spectra. This is especially true of the spectra of chemically peculiar (Ap) stars. This is clearly visible in the right panel of Fig.~\ref{fig:hfs}, which compares observed and synthetic profiles for the Lu~II~$\lambda$5983.86~\AA\ line in the spectrum of one of the most peculiar Ap-stars, Przybylski's star (HD~101065). The observed spectral line profile has a complex structure, in which individual HFS components are partially resolved, which cannot be described without allowance for HFS.

\section{CONCLUSION}
\noindent

We have created a database of HFS constants for VALD, which can be used to describe HFS in neutral and singly-ionized atoms. The database is sufficiently complete for analyses of spectra of A--M stars in the near UV, optical, and IR. Studies of the spectra of hot O--B stars require the use of laboratory measurements for atoms in the second and higher stages of ionization.

%

\bibliographystyle{mnras}
\bibliography{paper}

\onecolumn
\thispagestyle{empty}
\begin{sidewaystable}
\caption{Number of observed and split lines with a given central depth for various isotopes in the spectra of O--K stars.\label{tab:nelem}}
\renewcommand\arraystretch{2}
\label{tab:nelem}
\tabcolsep=1.5mm
\begin{tabular}{|c|r|r|r|r||r|r|r|r||r|r|r|r||r|r|r|r||r|r|r|r||r|r|r|r||r|r|r|r|}
\hline
                    &             \multicolumn{4}{c||}{O}                   &             \multicolumn{4}{c||}{B}                   &             \multicolumn{4}{c||}{A}                   &             \multicolumn{4}{c||}{F}                   &             \multicolumn{4}{c||}{G}                   &             \multicolumn{4}{c||}{K}                   &             \multicolumn{4}{c|}{M}              \\
&\multicolumn{2}{c|}{<.015}&\multicolumn{2}{c||}{<.300}&\multicolumn{2}{c|}{<.015}&\multicolumn{2}{c||}{<.300}&\multicolumn{2}{c|}{<.015}&\multicolumn{2}{c||}{<.300}&\multicolumn{2}{c|}{<.015}&\multicolumn{2}{c||}{<.300}&\multicolumn{2}{c|}{<.015}&\multicolumn{2}{c||}{<.300}&\multicolumn{2}{c|}{<.015}&\multicolumn{2}{c||}{<.300}&\multicolumn{2}{c|}{<.015}&\multicolumn{2}{c|}{<.300}\\
\hline
$^{6}$\ion{Li}{1}   &           --&           --&           --&           --&           --&           --&           --&           --&           --&           --&           --&           --&           --&           --&           --&           --&           --&           --&           --&           --&           --&           --&           --&           --&            1&            1&           --&           --\\
$^{7}$\ion{Li}{1}   &           --&           --&           --&           --&           --&           --&           --&           --&           --&           --&           --&           --&           --&           --&           --&           --&           --&           --&           --&           --&            2&            2&           --&           --&            3&            2&            2&            2\\
$^{23}$\ion{Na}{1}  &           --&           --&           --&           --&            1&            1&           --&           --&           13&           11&            1&            1&           50&           40&            5&            5&           69&           51&           11&           11&          108&           78&           32&           30&          129&           91&           48&           40\\
$^{27}$\ion{Al}{1}  &           --&           --&           --&           --&           --&           --&           --&           --&           66&           11&           17&            5&          188&           18&           70&            9&          215&           22&          114&            9&          256&           36&          129&           11&          208&           32&           62&           11\\
$^{27}$\ion{Al}{2}  &          ~~2&          ~~1&         ~~--&         ~~--&           94&           22&           23&           11&           57&           14&           24&           10&           27&            9&           16&            7&           18&            7&           14&            7&           10&            4&            2&           --&            2&           --&            2&           --\\
$^{39}$\ion{K}{1}   &           --&           --&           --&           --&           --&           --&           --&           --&            1&            1&           --&           --&            8&            5&            1&            1&           11&            7&            1&            1&           23&           15&            4&            3&           50&           20&            8&            6\\
$^{40}$\ion{K}{1}   &           --&           --&           --&           --&           --&           --&           --&           --&           --&           --&           --&           --&           --&           --&           --&           --&           --&           --&           --&           --&           --&           --&           --&           --&           --&           --&           --&           --\\
$^{41}$\ion{K}{1}   &           --&           --&           --&           --&           --&           --&           --&           --&           --&           --&           --&           --&           --&           --&           --&           --&           --&           --&           --&           --&           --&           --&           --&           --&           --&           --&           --&           --\\
$^{45}$\ion{Sc}{1}  &           --&           --&           --&           --&           --&           --&           --&           --&            1&            1&           --&           --&           15&            4&           --&           --&           34&           10&            5&            3&          207&           35&           46&           13&          332&           49&          100&           24\\
$^{45}$\ion{Sc}{2}  &           --&           --&           --&           --&           --&           --&           --&           --&           91&           35&           18&           10&          103&           48&           27&           24&           89&           46&           30&           27&           68&           45&           31&           29&           37&           34&           23&           22\\
$^{47}$\ion{Ti}{1}  &           --&           --&           --&           --&           --&           --&           --&           --&           --&           --&           --&           --&           --&           --&           --&           --&           --&           --&           --&           --&           --&           --&           --&           --&           --&           --&           --&           --\\
$^{47}$\ion{Ti}{2}  &           --&           --&           --&           --&           --&           --&           --&           --&           --&           --&           --&           --&           --&           --&           --&           --&           --&           --&           --&           --&           --&           --&           --&           --&           --&           --&           --&           --\\
$^{49}$\ion{Ti}{1}  &           --&           --&           --&           --&           --&           --&           --&           --&           --&           --&           --&           --&           --&           --&           --&           --&           --&           --&           --&           --&           --&           --&           --&           --&           --&           --&           --&           --\\
$^{49}$\ion{Ti}{2}  &           --&           --&           --&           --&           --&           --&           --&           --&           --&           --&           --&           --&           --&           --&           --&           --&           --&           --&           --&           --&           --&           --&           --&           --&           --&           --&           --&           --\\
$^{50}$\ion{V}{1}   &           --&           --&           --&           --&           --&           --&           --&           --&           --&           --&           --&           --&           --&           --&           --&           --&           --&           --&           --&           --&           --&           --&           --&           --&           --&           --&           --&           --\\
$^{51}$\ion{V}{1}   &           --&           --&           --&           --&           --&           --&           --&           --&            3&            3&           --&           --&          115&           91&            7&            7&          292&          191&           31&           30&         1271&          775&          426&          288&         1555&          958&          689&          459\\
$^{51}$\ion{V}{2}   &           --&           --&           --&           --&            1&           --&           --&           --&          546&          128&           87&           37&          796&          200&          256&           88&          791&          211&          291&          109&          519&          177&          231&          102&          237&          106&           81&           44\\
$^{55}$\ion{Mn}{1}  &           --&           --&           --&           --&           --&           --&           --&           --&           60&           41&            5&            5&          437&          146&           66&           39&          649&          187&          177&           75&         1092&          271&          429&          143&          760&          223&          325&          120\\
$^{55}$\ion{Mn}{2}  &           --&           --&           --&           --&          177&           16&            4&           --&         1466&           50&          398&           25&         1430&           51&          507&           32&         1243&           50&          476&           33&          429&           30&          144&           23&           43&           11&           12&           --\\
\hline
\end{tabular}
\end{sidewaystable}

\clearpage
\thispagestyle{empty}
\begin{sidewaystable}[t]
	\centering
\addtocounter{table}{-1}
\caption{Number of observed and split lines with a given central depth for various isotopes in the spectra of O--K stars (continue).}
\renewcommand\arraystretch{2}
\tabcolsep=1.5mm
\begin{tabular}{|c|r|r|r|r||r|r|r|r||r|r|r|r||r|r|r|r||r|r|r|r||r|r|r|r||r|r|r|r|}
\hline
                    &             \multicolumn{4}{c||}{O}                   &             \multicolumn{4}{c||}{B}                   &             \multicolumn{4}{c||}{A}                   &             \multicolumn{4}{c||}{F}                   &             \multicolumn{4}{c||}{G}                   &             \multicolumn{4}{c||}{K}                   &             \multicolumn{4}{c|}{M}              \\
&\multicolumn{2}{c|}{<.015}&\multicolumn{2}{c||}{<.300}&\multicolumn{2}{c|}{<.015}&\multicolumn{2}{c||}{<.300}&\multicolumn{2}{c|}{<.015}&\multicolumn{2}{c||}{<.300}&\multicolumn{2}{c|}{<.015}&\multicolumn{2}{c||}{<.300}&\multicolumn{2}{c|}{<.015}&\multicolumn{2}{c||}{<.300}&\multicolumn{2}{c|}{<.015}&\multicolumn{2}{c||}{<.300}&\multicolumn{2}{c|}{<.015}&\multicolumn{2}{c|}{<.300}\\
\hline
$^{57}$\ion{Fe}{1}  &         ~~--&         ~~--&         ~~--&         ~~--&         ~~--&         ~~--&         ~~--&         ~~--&           --&           --&           --&           --&           --&           --&           --&           --&           --&           --&           --&           --&           --&           --&           --&           --&           --&           --&           --&           --\\
$^{59}$\ion{Co}{1}  &           --&           --&           --&           --&           --&           --&           --&           --&           74&           74&            2&            2&          597&          595&          142&          142&          967&          963&          322&          321&         1463&         1457&          722&          720&         1144&         1141&          616&          615\\
$^{59}$\ion{Co}{2}  &           --&           --&           --&           --&          146&            2&            2&           --&          556&            2&          206&            2&          628&            2&          295&            2&          569&            2&          300&            2&          221&            2&           92&            2&           43&            2&           19&           --\\
$^{61}$\ion{Ni}{1}  &           --&           --&           --&           --&           --&           --&           --&           --&           --&           --&           --&           --&           --&           --&           --&           --&           --&           --&           --&           --&           --&           --&           --&           --&           --&           --&           --&           --\\
$^{63}$\ion{Cu}{1}  &           --&           --&           --&           --&           --&           --&           --&           --&           43&           21&            1&            1&           89&           66&           27&           24&          115&           75&           43&           39&          174&           96&           72&           56&          181&           73&          140&           56\\
$^{65}$\ion{Cu}{1}  &           --&           --&           --&           --&           --&           --&           --&           --&            2&            2&           --&           --&           17&           17&            2&            2&           22&           22&            8&            8&           29&           29&           16&           16&           21&           21&           14&           14\\
$^{67}$\ion{Zn}{1}  &           --&           --&           --&           --&           --&           --&           --&           --&           --&           --&           --&           --&           --&           --&           --&           --&           --&           --&           --&           --&           --&           --&           --&           --&           --&           --&           --&           --\\
$^{67}$\ion{Zn}{2}  &           --&           --&           --&           --&           --&           --&           --&           --&           --&           --&           --&           --&           --&           --&           --&           --&           --&           --&           --&           --&           --&           --&           --&           --&           --&           --&           --&           --\\
$^{69}$\ion{Ga}{1}  &           --&           --&           --&           --&           --&           --&           --&           --&            1&           --&           --&           --&            4&           --&           --&           --&            6&           --&            2&           --&           12&           --&            8&           --&           13&           --&           11&           --\\
$^{69}$\ion{Ga}{2}  &           --&           --&           --&           --&            3&           --&            1&           --&            4&           --&            1&           --&            2&           --&            1&           --&            1&           --&            1&           --&            1&           --&            1&           --&            1&           --&           --&           --\\
$^{71}$\ion{Ga}{1}  &           --&           --&           --&           --&           --&           --&           --&           --&           --&           --&           --&           --&           --&           --&           --&           --&           --&           --&           --&           --&           --&           --&           --&           --&           --&           --&           --&           --\\
$^{71}$\ion{Ga}{2}  &           --&           --&           --&           --&           --&           --&           --&           --&           --&           --&           --&           --&           --&           --&           --&           --&           --&           --&           --&           --&           --&           --&           --&           --&           --&           --&           --&           --\\
$^{85}$\ion{Rb}{1}  &           --&           --&           --&           --&           --&           --&           --&           --&           --&           --&           --&           --&           --&           --&           --&           --&           --&           --&           --&           --&            2&            1&           --&           --&            5&            1&            1&            1\\
$^{87}$\ion{Rb}{1}  &           --&           --&           --&           --&           --&           --&           --&           --&           --&           --&           --&           --&           --&           --&           --&           --&           --&           --&           --&           --&           --&           --&           --&           --&           --&           --&           --&           --\\
$^{89}$\ion{Y}{2}   &           --&           --&           --&           --&           --&           --&           --&           --&           51&           20&            4&            3&           82&           31&           20&           15&           77&           34&           23&           19&           49&           34&           23&           21&           24&           22&           14&           13\\
$^{93}$\ion{Nb}{2}  &           --&           --&           --&           --&           --&           --&           --&           --&            7&            5&           --&           --&           69&           42&            3&            3&          125&           58&           13&           12&          112&           60&           25&           23&           38&           36&            1&            1\\
$^{95}$\ion{Mo}{2}  &           --&           --&           --&           --&           --&           --&           --&           --&           --&           --&           --&           --&           --&           --&           --&           --&           --&           --&           --&           --&           --&           --&           --&           --&           --&           --&           --&           --\\
$^{97}$\ion{Mo}{2}  &           --&           --&           --&           --&           --&           --&           --&           --&           --&           --&           --&           --&           --&           --&           --&           --&           --&           --&           --&           --&           --&           --&           --&           --&           --&           --&           --&           --\\
$^{127}$\ion{I}{1}  &           --&           --&           --&           --&           --&           --&           --&           --&           --&           --&           --&           --&           --&           --&           --&           --&           --&           --&           --&           --&           --&           --&           --&           --&           --&           --&           --&           --\\
$^{127}$\ion{I}{2}  &           --&           --&           --&           --&           --&           --&           --&           --&           --&           --&           --&           --&           --&           --&           --&           --&           --&           --&           --&           --&           --&           --&           --&           --&           --&           --&           --&           --\\
\hline
\end{tabular}
\end{sidewaystable}

\clearpage
\thispagestyle{empty}
\begin{sidewaystable}[t]
	\centering
\addtocounter{table}{-1}
\caption{Number of observed and split lines with a given central depth for various isotopes in the spectra of O--K stars (continue).}
\renewcommand\arraystretch{2}
\tabcolsep=1.5mm
\begin{tabular}{|c|r|r|r|r||r|r|r|r||r|r|r|r||r|r|r|r||r|r|r|r||r|r|r|r||r|r|r|r|}
\hline
                    &             \multicolumn{4}{c||}{O}                   &             \multicolumn{4}{c||}{B}                   &             \multicolumn{4}{c||}{A}                   &             \multicolumn{4}{c||}{F}                   &             \multicolumn{4}{c||}{G}                   &             \multicolumn{4}{c||}{K}                   &             \multicolumn{4}{c|}{M}              \\
&\multicolumn{2}{c|}{<.015}&\multicolumn{2}{c||}{<.300}&\multicolumn{2}{c|}{<.015}&\multicolumn{2}{c||}{<.300}&\multicolumn{2}{c|}{<.015}&\multicolumn{2}{c||}{<.300}&\multicolumn{2}{c|}{<.015}&\multicolumn{2}{c||}{<.300}&\multicolumn{2}{c|}{<.015}&\multicolumn{2}{c||}{<.300}&\multicolumn{2}{c|}{<.015}&\multicolumn{2}{c||}{<.300}&\multicolumn{2}{c|}{<.015}&\multicolumn{2}{c|}{<.300}\\
\hline
$^{135}$\ion{Ba}{2} &         ~~--&         ~~--&         ~~--&         ~~--&         ~~--&         ~~--&         ~~--&         ~~--&         ~~--&         ~~--&         ~~--&         ~~--&          ~~2&          ~~2&         ~~--&         ~~--&          ~~2&          ~~2&          ~~1&          ~~1&          ~~2&          ~~2&          ~~1&          ~~1&          ~~2&          ~~2&          ~~1&          ~~1\\
$^{137}$\ion{Ba}{2} &           --&           --&           --&           --&           --&           --&           --&           --&            1&            1&           --&           --&            1&            1&            1&            1&            1&            1&            1&            1&            1&            1&            1&            1&            1&            1&            1&            1\\
$^{139}$\ion{La}{2} &           --&           --&           --&           --&           --&           --&           --&           --&            5&            5&           --&           --&           53&           53&            5&            5&           66&           66&            9&            9&           78&           78&           16&           16&           48&           48&           11&           11\\
$^{141}$\ion{Pr}{1} &           --&           --&           --&           --&           --&           --&           --&           --&           --&           --&           --&           --&           --&           --&           --&           --&           --&           --&           --&           --&           --&           --&           --&           --&           29&           --&           --&           --\\
$^{141}$\ion{Pr}{2} &           --&           --&           --&           --&           --&           --&           --&           --&            2&            2&           --&           --&           36&           36&           --&           --&           44&           44&           --&           --&           65&           64&            5&            5&           79&           78&           13&           13\\
$^{151}$\ion{Eu}{2} &           --&           --&           --&           --&           --&           --&           --&           --&            1&            1&           --&           --&            5&            5&            1&            1&            6&            6&            1&            1&            6&            6&            3&            3&            5&            5&            3&            3\\
$^{153}$\ion{Eu}{2} &           --&           --&           --&           --&           --&           --&           --&           --&            1&            1&           --&           --&            6&            6&            1&            1&            6&            6&            3&            3&            6&            6&            4&            4&            6&            6&            4&            4\\
$^{155}$\ion{Gd}{2} &           --&           --&           --&           --&           --&           --&           --&           --&           --&           --&           --&           --&           --&           --&           --&           --&           --&           --&           --&           --&           --&           --&           --&           --&           --&           --&           --&           --\\
$^{157}$\ion{Gd}{2} &           --&           --&           --&           --&           --&           --&           --&           --&           --&           --&           --&           --&           --&           --&           --&           --&           --&           --&           --&           --&           --&           --&           --&           --&           --&           --&           --&           --\\
$^{159}$\ion{Tb}{2} &           --&           --&           --&           --&           --&           --&           --&           --&            1&            1&           --&           --&           18&           15&           --&           --&           32&           29&           --&           --&           61&           53&            3&            3&           47&           40&            4&            4\\
$^{165}$\ion{Ho}{1} &           --&           --&           --&           --&           --&           --&           --&           --&           --&           --&           --&           --&           --&           --&           --&           --&           --&           --&           --&           --&            2&            1&           --&           --&           11&            5&            2&            1\\
$^{175}$\ion{Lu}{1} &           --&           --&           --&           --&           --&           --&           --&           --&           --&           --&           --&           --&           --&           --&           --&           --&           --&           --&           --&           --&            4&           --&           --&           --&           22&            4&            5&           --\\
$^{175}$\ion{Lu}{2} &           --&           --&           --&           --&           --&           --&           --&           --&           --&           --&           --&           --&            6&            1&           --&           --&           11&            1&           --&           --&           11&            1&            1&            1&            3&            1&            1&            1\\
$^{176}$\ion{Lu}{1} &           --&           --&           --&           --&           --&           --&           --&           --&           --&           --&           --&           --&           --&           --&           --&           --&           --&           --&           --&           --&           --&           --&           --&           --&           --&           --&           --&           --\\
$^{181}$\ion{Ta}{1} &           --&           --&           --&           --&           --&           --&           --&           --&           --&           --&           --&           --&           --&           --&           --&           --&           --&           --&           --&           --&           26&           26&           --&           --&           41&           41&            5&            5\\
$^{181}$\ion{Ta}{2} &           --&           --&           --&           --&           --&           --&           --&           --&            1&            1&           --&           --&           38&           38&           --&           --&           79&           73&            1&            1&           41&           38&            1&            1&           --&           --&           --&           --\\
$^{203}$\ion{Tl}{1} &           --&           --&           --&           --&           --&           --&           --&           --&           --&           --&           --&           --&           --&           --&           --&           --&           --&           --&           --&           --&           --&           --&           --&           --&           --&           --&           --&           --\\
$^{205}$\ion{Tl}{1} &           --&           --&           --&           --&           --&           --&           --&           --&           --&           --&           --&           --&           --&           --&           --&           --&           --&           --&           --&           --&            5&            1&            1&            1&            7&            1&            4&            1\\
$^{209}$\ion{Bi}{3} &           --&           --&           --&           --&           --&           --&           --&           --&           --&           --&           --&           --&           --&           --&           --&           --&           --&           --&           --&           --&           --&           --&           --&           --&           --&           --&           --&           --\\
\hline
\end{tabular}
\end{sidewaystable}

\end{document}